\documentclass[aps,12pt]{revtex4}
\usepackage[english]{babel}
\usepackage{amsmath,amssymb}

\newcommand{\A}{\mathcal{A}}
\newcommand{\Fn}{\mathrm{Fun}(\R^2)}
\newcommand{\Fns}{\mathrm{Fun}^\infty(\R^2)}
\newcommand{\B}{\mathcal{B}}
\newcommand{\R}{\mathbb{R}}
\newcommand{\C}{\mathbb{C}}
\newcommand{\U}{\mathcal{U}}
\newcommand{\K}{\mathcal{K}}
\newcommand{\HH}{\mathcal{H}}
\newcommand{\V}{\mathcal{V}}
\newcommand{\g}{\mathfrak{g}}
\newcommand{\de}{\mathrm{d}}
\newcommand{\ma}[1]{\small\Big(\!\begin{array}{cc}#1\end{array}\!\Big)}
\newcommand{\az}{\triangleright}
\newcommand{\inner}[1]{\left<#1\right>}
\newcommand{\Da}[1]{#1\otimes 1+1\otimes #1}
\newcommand{\dt}{\,\dot{\otimes}\,}
\newcommand{\bn}[1]{\textrm{\small$\Big(\!\begin{array}{c} #1 \end{array}\!\Big)$}}

\newcommand{\bcp}{\vartriangleright\!\blacktriangleleft}
\newcommand{\D}{D\mkern-11.5mu/\,}
\newcommand{\nint}{\int\mkern-19mu-\;}
\newcommand{\tr}{\mathrm{Trace}}

\newtheorem{df}{Definition}
\newtheorem{prop}[df]{Proposition}
\newtheorem{cor}[df]{Corollary}

\newcommand{\proof}[1]{\noindent\textit{Proof:}\vspace{5pt} \\ 
   \parbox{\textwidth}{#1\hfill$\square$} }

\usepackage[bookmarksopen=true]{hyperref}
\hypersetup{%
    pdftitle={Spectral geometry of k-Minkowski space},
    pdfauthor={Francesco D'Andrea},
    pdfsubject={Noncommutative geometry and quantum groups},
    colorlinks=true,
    pdfpagemode=UseOutlines,
    pdfstartpage=1,
    pdfstartview=FitH,
    bookmarksopen=false,
    linkcolor=red,
    anchorcolor=black,
    citecolor=blue,
    urlcolor=cyan}

\begin{document}

\title{\vspace*{1cm}Spectral geometry of $\kappa$-Minkowski space}

\author{\vspace{5mm}Francesco~D'ANDREA\vspace{3pt}}
 \affiliation{International School for Advanced Studies (SISSA), I--34014, Trieste,
 Italy\vspace{3pt}}

\begin{abstract}\vspace{5mm}\noindent%
After recalling Snyder's idea~\cite{Sny47-QST} of using vector fields
over a smooth manifold as `coordinates on a noncommutative space',
we discuss a two dimensional toy-model whose `dual'
noncommutative coordinates form a Lie algebra: this is the well
known $\kappa$-Minkowski space~\cite{Maj94-Bcp}.

We show how to improve Snyder's idea using the tools of quantum
groups and noncommutative geometry.

We find a natural representation of the coordinate algebra of $\kappa$-Minkowski
as linear operators on an Hilbert space (a major problem in the construction
of a physical theory), study its `spectral properties' and discuss how
to obtain a Dirac operator for this space.

We describe two Dirac operators.

The first is associated with a spectral triple.
We prove that the cyclic integral of M.~Dimitrijevic et
al.~\cite{DJMTWW} can be obtained as Dixmier trace
associated to this triple.

The second Dirac operator is equivariant for the action of the quantum
Euclidean group, but it has unbounded commutators with the algebra.
\end{abstract}

\maketitle

\hfill\eject

\section{Introduction}\noindent
In classical mechanics, Legendre transformation in the Hamiltonian formalism
exhibits an evident symmetry between coordinate space and momentum space.

The same holds in quantum mechanics on a flat space ($\R^n$ or a quotient
by a discrete subgroup), with Fourier transform/series replacing the
Legendre transformation.

Dealing with quantum mechanics on a curved space $M$ (a smooth manifold),
the situation changes. While coordinate functions on $M$ are elements of
a commutative algebra, `momenta' (we mean, vector fields over $M$) no
longer commute. A basic example is the $3$-sphere, whose vector fields
generate $su(2)$, the algebra of angular momenta of quantum mechanics.

Snyder's idea was to use `momenta', i.e.~vector fields over a smooth
manifold $M$, as ``coordinates on a noncommutative space'' dual to $M$.

He applied this idea to the four-dimensional deSitter space, and was led to study
the algebra generated by ten operators $\{x_\mu,J_{\mu\nu}\}$ satisfying the
commutation rules (Greek letters run over $0,1,2,3$):
\begin{equation}\label{eq:sny}
[x_\mu,x_\nu]=i\lambda^2J_{\mu\nu}\;\;,
\end{equation}
where $J_{\mu\nu}$ is an element of $so(3,1)$ with suitable commutators
with the ``coordinates'' $x_\mu$.
One can notice that this algebra (isomorphic to $so(3,2)$) has too much
generators to represent a four-dimensional `noncommutative space', while
the elements $x_\mu$ alone are correct in number, but don't close an algebra.

A characteristic feature is the presence in (\ref{eq:sny}) of an invariant
length $\lambda$, whose role is to provide the correct physical dimensions and
to allow us to recover $\R^4$ as $\lambda$ goes to zero (`long distances'
or `low energy' limit).

The purpose of this notes is to show how to modify Snyder's idea in
such a way to obtain an algebra that can be studied with
noncommutative-geometric tools.

In sec.~\ref{sec:due}, we discuss Snyder's model in full detail in the two dimensional
version, and realize that Snyder's coordinates don't close an algebra, neither
in two dimensions, nor in four.

In sec.~\ref{sec:tre}, we retry in three dimensions, and realize a fundamental difference
with the original Snyder example: in the three dimensional case the `dual' of
the deSitter space is an algebra. This is a basic property, if we want to study
noncommutative geometry~\cite{ConBook,ConRe,Con96-Gra}.

The geometrical reason at the base of the difference between Snyder's model in
three and in $n\neq 3$ dimensions is clear: the deSitter space in three dimensions
is a Lie group, hence the tangent bundle trivialize and vector fields form
a Lie algebra isomorphic to the Lie algebra of the group.

In sec.~\ref{sec:quattro}, we search for a two-dimensional example and realize
that there is only one, the spacetime studied in literature under the name of
$\kappa$-Minkowski~\cite{Maj94-Bcp}, whose coordinate algebra is isomorphic to
$U(sb(2,\R))$.

As a byproduct of this section, we find a natural representation
of the algebra as linear operators on a Hilbert space (the choice
of a representation on a Hilbert space is a major problem in the
construction of a physical theory).
Another main point of this section is the realization of Poincar\'e as
a symmetry of $\kappa$-Minkowski, in the spirit of Snyder's idea.

In sec.~\ref{sec:cinque} we adopt a different point of view and,
applying Weyl quantization, construct a $C^*$-algebra
encoding informations on the `topology' of the underlying
$\kappa$-Minkowski space.

We then focus the attention on the construction of a spectral triple
over $\kappa$-Minkowski space. A proposal for a Dirac operator on
$\kappa$-Minkowski already appeared in~\cite{NowD}, although the Dirac
operator they found doesn't satisfy Connes axioms for a spectral triple
(the same holds for the one parametric family of Dirac operators
in~\cite{Bib}). For the generalization of such axioms to the
non-compact case we refer to the original paper of A.~Connes~\cite{ConRe}
(see also~\cite{Lizzi,Gay03} for a comprehensive presentation).

In section \ref{sec:Dirac}, we construct a Dirac operator that fulfil
all axioms of a spectral triple, and prove that the Dixmier trace
associated to the spectral triple is just the cyclic integral
discussed in~\cite{DJMTWW}.

Then, we compute a Dirac operator imposing equivariance for the action
of the quantum Euclidean group. This turns out to be very similar
to the one in~\cite{NowD}, and has not bounded commutators with the
algebra.

\section{The two-dimensional analogue of Snyder's spacetime}\label{sec:due}\noindent
In his original article, Snyder considers the deSitter space $SO(3,2)/SO(3,1)$ and
identifies the spacetime ``coordinate functions'' with a basis of the vector
subspace of $so(3,2)$ orthogonal to $so(3,1)$. The (connected component of
the) $SO(3,1)$ subgroup provides the Lorentz symmetries and, with a suitable
choice of the momenta, it can be extended to a Poincar\'{e} symmetry.
For the sake of simplicity, we will study, in full details, the two-dimensional
version. The interested readers can find the four-dimensional model discussed
in Snyder's original article~\cite{Sny47-QST}.

In the two-dimensional version, the deSitter space is $SO(2,1)/SO(1,1)$.
If $\eta_0,\eta_1,\eta_2$ are coordinates of $\R^3$, the deSitter spacetime
is the $SO(2,1)$-orbit of equation
\begin{equation}\label{eq:deSitter}
\eta_0^2-\eta_1^2-\eta_2^2=-1\;\;.
\end{equation}
The (three dimensional Lorentz) Lie algebra $so(2,1)$ has generators
$\{J,K_1,K_2\}$ (a rotation and two boosts) given by
\begin{align*}
 & J=i(\eta_2\partial_1-\eta_1\partial_2)   \;\;, \\
 & K_1=i(\eta_0\partial_1+\eta_1\partial_0) \;\;, \\
 & K_2=i(\eta_0\partial_2+\eta_2\partial_0) \;\;,
\end{align*}
where $\partial_\mu=\partial/\partial\eta_\mu$. So, the commutation rules are:
\begin{equation}\label{eq:Sny2}
[J,K_1]=iK_2\;\;,\qquad [J,K_2]=-iK_1\;\;,\qquad [K_1,K_2]=-iJ\;\;.
\end{equation}
Emulating Snyder, we call $N=-K_1$ the generator of the $so(1,1)$ subalgebra (the
one-dimensional Lorentz Lie-algebra, isomorphic to $\R$) that leave fixed the
point $m_0=(0,0,1)$, and $x=\lambda J$, $t=\lambda K_1$ the remaining generators
(whose span is the subspace of vector fields over $SO(2,1)$ that are tangent
to $M$ at $m_0$).
The length $\lambda$ provides the correct physical dimensions. By (\ref{eq:Sny2}),
the coordinates satisfy the commutation rule
\begin{equation*}
[x,t]=i\lambda^2N\;\;,
\end{equation*}
and clearly don't form a subalgebra of $so(2,1)$.

The higher-dimensional analogue is the commutator $[x_\mu,x_\nu]=i\lambda^2J_{\mu\nu}$
anticipated in the introduction.

The action of the Lorentz algebra $so(1,1)$ on the coordinates is via
commutator, and by (\ref{eq:Sny2}) is undeformed:
\begin{equation*}
N\az x=it\;\;,\qquad N\az t=ix\;\;.
\end{equation*}
Since the action is undeformed, the invariant is the classical quadratic
element $x^2-t^2$.

We need to introduce translations, yet.

Let us recall the idea of Snyder: to start with a commutative spacetime, in
which translations do not commute, and define a noncommutative spacetime with
commutative momenta. Half of this idea was already applied, to find the
noncommutative spacetime ``dual'' to deSitter.

To complete the picture, followind Snyder, we define ``momenta'' as
$P=\lambda^{-1}\eta_1$ and $E=\lambda^{-1}\eta_0$ (the presence of $\lambda$
ensures the correct physical dimensions). Together with $N$ they form the
classical Poincar\'{e} algebra:
\begin{equation*}
[E,P]=0\;\;,\qquad [N,P]=-iE\;\;,\qquad [N,E]=-iP\;\;.
\end{equation*}

The ``momenta'' act on the coordinates via commutators. For example:
\begin{equation*}
P\az x=[P,x]=-i\lambda(\eta_2\partial_1-\eta_1\partial_2)\lambda^{-1}\eta_1
      =-i\eta_2=-i\sqrt{\lambda^2(E^2-P^2)-1}\;\;.
\end{equation*}
In the last step we used the equation (\ref{eq:deSitter}), defining the
deSitter space.

With a straightforward calculation we derive the (deformed) action of momenta
on coordinates
\begin{equation*}
P\az x=-i\sqrt{\lambda^2(E^2-P^2)-1}=E\az t\;\;, \qquad P\az t=E\az x=0\;\;,
\end{equation*}
while the associated phase-space is defined by
\begin{equation*}
[x,P]=i\sqrt{\lambda^2(E^2-P^2)-1}=[t,E]\;\;, \qquad [t,P]=[x,E]=0\;\;.
\end{equation*}
In the $\lambda\to 0$ limit, the spacetime reduces to a commutative one,
phase-space becomes the Heisenberg algebra and the action of the Poincar\'{e}
algebra reduces to the standard one; so for $\lambda=0$ we recover the
classical scenario.

\section{The three-dimensional analogue of Snyder's spacetime}\label{sec:tre}\noindent
In two dimensions, we have just seen that ``Snyder's coordinates'' don't form a
complete set of generators for an algebra. Let us try in three dimensions.

The space we consider here is the $SO(2,2)$-orbit $M\subset\R^4$ of equation:
\begin{equation}\label{eq:deS3}
\eta_0^2+\eta_1^2-\eta_2^2-\eta_3^2=-1\;\;,\qquad\eta_\mu\in\R^4\;.
\end{equation}
The stability group of the point $m_0=(0,0,0,1)$ is $SO(2,1)$, so the orbit is the
quotient $M\simeq SO(2,2)/SO(2,1)$. Being an homogeneous $SO(2,2)$-space,
elements of the Lie algebra $so(2,2)$ (i.e., vector fields over $SO(2,2)$)
correspond to derivatives of $C^\infty(M)$. Vectors of the subspace
$so(2,1)\subset so(2,2)$ are orthogonal to $M$, so the ``naive'' approach would
be to take a basis of $so(2,1)^\perp$, the orthogonal of $so(2,1)$ in $so(2,2)$,
as coordinates of the noncommutative spacetime dual to $M$. But $so(2,1)^\perp$ is
not a subalgebra of $so(2,2)$, this because $M$ is not a group with the
quotient structure ($SO(2,1)$ is not a \emph{normal} subgroup of $SO(2,2)$).

Despite this, we can put a Lie group structure on $M$. Writing
\begin{equation}\label{eq:g}
g=\ma{\eta_0+\eta_3 & \eta_1+\eta_2 \\ \eta_1-\eta_2 & -\eta_0+\eta_3}
\;\;,\qquad\eta_\mu\in\R^4\;,
\end{equation}
we have the obvious (smooth manifold) isomorphism $M\simeq SL(2,\R)$, the
equation $\det g=1$ (that identifies $SL(2,\R)$ inside $\mathrm{Mat}(2,\R)$)
being equivalent to (\ref{eq:deS3}). Thus, $M$ is a double cover of the Lorentz
group $SO(2,1)$.

This is  the pseudo-Euclidean analogue of the fibration
$SO(4)\!\stackrel{\textrm{\tiny{$SO(3)$}}}{\longrightarrow}\!S^3\simeq SU(2)$, i.e.\
$SO(2,2)\!\stackrel{\textrm{\tiny{$SO(2,1)$}}}{\longrightarrow}\!SL(2,\R)$.

Since $M$ is a Lie group, its tangent bundle is trivial. The Lie algebra of
global vector fields on $M$, isomorphic to $sl(2,\R)$, will be identified with
the algebra ``coordinate functions on the noncommutative spacetime'', three
dimensional analogue of the Snyder's spacetime. Let us compute it explicitly.

We take $\tilde{L}\in sl(2,\R)$ (real traceless matrices) and call $L$ the
associated vector field on $M$, defined by:
\begin{equation}\label{eq:vf}
(Lf)(g)=\frac{\de}{\de\tau}\Big|_{\tau=0}f(\exp\{\tau L\}\cdot g)\;\;,
\end{equation}
for all $f\in C^\infty(M)$, $g\in SL(2,\R)$. We fix a basis for $sl(2,\R)$:
\begin{equation*}
 \tilde{t}=\ma{1 & 0 \\ 0 & -1}\;\;,\qquad
 \tilde{x}=\ma{0 & 1 \\ 1 & 0}\;\;,\qquad
 \tilde{y}=\ma{0 & 1 \\ -1 & 0}\;\;,
\end{equation*}
compute the exponentials
\begin{align*}
 \exp\{\tau\tilde{t}\} &=\ma{e^{\tau} & 0 \\ 0 & e^{-\tau}} \;\;, \\
 \exp\{\tau\tilde{x}\} &=\ma{\cosh\tau & \sinh\tau \\ \sinh\tau & \cosh\tau} \;\;, \\
 \exp\{\tau\tilde{y}\} &=\ma{\cos\tau & \sin\tau \\ -\sin\tau & \cos\tau} \;\;,
\end{align*}
and their action via left multiplication on (\ref{eq:g}), the generic element
of $SL(2,\R)$. Then, through (\ref{eq:vf}), we determine the associated vector
fields. For example:
\begin{equation*}
\exp\{\tau\tilde{t}\}g=\left(\!\begin{array}{cc} (\eta_0+\eta_3)e^{\tau} &
(\eta_1+\eta_2)e^{\tau} \\ (\eta_1-\eta_2)e^{-\tau} & (-\eta_0+\eta_3)e^{-\tau}
\end{array}\!\right) \;\;.
\end{equation*}
So $\exp\{\tau\tilde{t}\}$ maps $\eta$ to the point $\eta^{(\tau)}$, with
coordinates
\begin{small}
\begin{center}
$\left(\!\begin{array}{c} \eta^{(\tau)}_0 \\ \eta^{(\tau)}_1 \\
\eta^{(\tau)}_2 \\ \eta^{(\tau)}_3
\end{array}\!\right)=\left(\!\begin{array}{c}
\eta_0\cosh\tau+\eta_3\sinh\tau \\ \eta_1\cosh\tau+\eta_2\sinh\tau \\
\eta_1\sinh\tau+\eta_2\cosh\tau \\ \eta_0\sinh\tau+\eta_3\cosh\tau
\end{array}\!\right)\quad${\normalsize ,}
\end{center}
\end{small}
and by (\ref{eq:vf}):
\begin{equation*}
(-i\lambda^{-1}t\cdot f)(\eta)=\frac{\de}{\de\tau}\Big|_{\tau=0}f(\eta^{(\tau)})\;\;,
\end{equation*}
where $\lambda$ is a parameter with the dimensions of a length.

We have called $-i\lambda^{-1}t$ the derivation associated to $\tilde{\tau}$
because we like to work with symmetric operators, and we want a $t$ with the
dimension of a length.

Thus,
\begin{equation*}
-i\lambda^{-1}t=\sum_{\mu}\frac{\de\eta^{(\tau)}_\mu}{\de\tau}\Big|_{\tau=0}\frac{\partial}
 {\partial\eta_\mu}=\eta_3\partial_0+\eta_2\partial_1+\eta_1\partial_2+\eta_0\partial_3 \;\;.
\end{equation*}
On the same line, one can compute $x$ and $y$. The full list of vector fields is:
\begin{align*}
 t &=i\lambda(
     \eta_3\partial_0+\eta_2\partial_1+\eta_1\partial_2+\eta_0\partial_3 ) \;\;, \\
 x &=i\lambda(
    -\eta_2\partial_0+\eta_3\partial_1-\eta_0\partial_2+\eta_1\partial_3 ) \;\;, \\
 y &=i\lambda(
     \eta_1\partial_0-\eta_0\partial_1+\eta_3\partial_2-\eta_2\partial_3 ) \;\;,
\end{align*}
and the commutation rules are those of ``$i\cdot sl(2,\R)$'':
\begin{equation*}
[t,x]=2i\lambda y\;\;,\qquad [t,y]=2i\lambda x\;\;,\qquad [x,y]=-2i\lambda t\;\;.
\end{equation*}
In contrast with the two dimensional Snyder's model, now $\{t,x,y\}$ is the basis of
a Lie algebra. Furthermore, this algebra has the correct number of generators
to represent a three-dimensional noncommutative space, dual to the three
dimensional deSitter space.

This idea can easily be generalized: starting with a Lie group $M$, one can
consider $U(\mathrm{Lie}\,M)$ as noncommutative space `dual' to $M$, and
eventually study it with tools of noncommutative geometry. 

A celebrated (compact, Riemannian) example is the \emph{fuzzy sphere}~\cite{Madore},
quotient of $U(su(2))$ for the ideal generated by $J^2-c$, with $J^2$ the Casimir
and $c$ a suitable constant.

In the next section we consider a simple, two-dimensional example.

\section{A two-dimensional model: $\kappa$-Minkowski}\label{sec:quattro}\noindent
In two dimensions, the unique (real connected) non-abelian Lie group
is the matrix group of elements~\cite[sec.~10.1]{Ful91-RT}
\begin{equation*}
\ma{a & b \\ 0 & 1}\;\;,\qquad(a,b)\in\R^+\!\times\R\;.
\end{equation*}
That is, (the connected component of) the group of affine transformations
of the real line:
\begin{equation*}
\bn{y \\ 1} \mapsto \ma{a & b \\ 0 & 1}\bn{y \\ 1}=\bn{ay+b \\ 1}\;\;,\qquad y\in\R\;.
\end{equation*}
The map
\begin{equation*}
\ma{a & b \\ 0 & 1}\mapsto\ma{a & a^{-1}b \\ 0 & a^{-1}}\;\;,
\end{equation*}
gives an isomorphism with $Sb(2,\R)$, the group of special upper-triangular
real matrices, so the variety $M\simeq Sb(2,\R)$ is the starting point.

We want to compute the vector fields and identify them with ``noncommutative
coordinates''. Let $L=\ma{a & b \\ 0 & 0}$ be the generic element of the
Lie algebra, $a,b\in \R$, and $-i\tilde{L}$ the associated vector field:
\begin{equation*}
(-i\tilde{L}f)(m)=\frac{\de}{\de\tau}\Big|_{\tau=0}f(\exp\{\tau L\}m)\;\;,
\qquad f\in C^\infty(M)\;\;.
\end{equation*}
We use physicists habit of working with selfadjoint operators. Since
\begin{equation*}
L^n=\ma{a^n & a^{n-1}b \\ 0 & 1}\;\;,\qquad
\exp\{\tau L\}=\ma{e^{\tau a} & \frac{e^{\tau a}-1}{a}b \\ 0 & 1} \;\;,
\end{equation*}
if we indicate with $m=\ma{\eta_0 & \eta_1 \\ 0 & 1}$ the generic point
of $M$, $(\eta_0,\eta_1)\in\R^+\!\times\R$,
it is easy to compute
\begin{equation*}
(\tilde{L}f)(\eta_0,\eta_1)=i\bigl(a\eta_0\partial_0+(a\eta_1+b)\partial_1\bigr)
f(\eta_0,\eta_1)\;\;,
\end{equation*}
with $\partial_\mu=\partial/\partial\eta_\mu$. We fix the basis
\begin{equation*}
x=i\lambda\partial_1\;\;,\qquad t=i\lambda(\eta_0\partial_0+\eta_1\partial_1)\;\;,
\end{equation*}
for vector fields on $M$, and define the algebra of ``functions on the noncommutative
spacetime'' to be the algebra of polynomials generated by $x$ and $t$.
The presence of the length $\lambda$ guarantees the correct
physical dimensions, and enables us to recover $\R^2$ as $\lambda\to 0$ limit.

Thus, the algebra is generated by $x$ and $t$ modulo
\begin{equation}\label{eq:kM}
[x,t]=i\lambda x\;\;.
\end{equation}
This is just $U(sb(2,\R))$, the two-dimensional version of the so-called
$\kappa$-Minkowski, introduced in~\cite{Maj94-Bcp} as homogeneous space
for $\kappa$-Poincar\'{e}~\cite{Luk91-qD}.

The same Lie-algebra, but with different real structure, emerges in the
Weyl quantization of $S^1\times\R$. 
In that case $x$ is unitary, thus the space is ``compact'' in the $x$-direction.

Now, let $\mu$ be the (left) Haar measure on $M$. Explicitly, for an integrable
function $f$ on $M$:
\begin{equation*}
\int_Mf\de\mu=\int_{\R^+\!\times\R}f(\eta_0,\eta_1)\eta_0^{-1}\de\eta_0\de\eta_1\;\;.
\end{equation*}
You can easily verify the invariance with respect to the left regular action,
i.e.~$\,\int_Mf\de\mu=\int_Mf'\de\mu\,$ with $\,f'(\eta_0,\eta_1)=f(a\eta_0,a\eta_1+b)\,$
and for all $\,(a,b)\in\R^+\!\times\R\,$.

With this measure, we can define the Hilbert space $\HH=\mathcal{L}^2(M,\mu)$
with inner product $\inner{\varphi,\psi}=\int_M\varphi^*\psi\de\,\mu$.

The measure being invariant, it means that `finite' transformations of the group
$M$ act as isometries on the associated Hilbert space, i.e.~as unitary
operators. Thus, the vector fields $x$ and $t$, the generators of these
transformations, are represented by (unbounded) selfadjoint linear
operators. I mean:
\begin{align*}
\inner{x^*\varphi,\psi} & :=\inner{\varphi,x\psi}\equiv\inner{x\varphi,\psi} \;\;, \\
\inner{t^*\varphi,\psi} & :=\inner{\varphi,t\psi}\equiv\inner{t\varphi,\psi} \;\;,
\end{align*}
for all $\varphi,\psi\in\HH$ and in the domain of $x$, resp.~$t$ (it is an easy
check to verify these equations). The self-adjointness of $x$ and $t$ allows us
to intepret them as quantum-mechanical observables.


\subsection{$\kappa$-Minkowski more in depth}\label{sec:sym}\noindent
Following Snyder's idea, we interpret vector fields on $M$ as ``coordinates'',
and define ``momenta'' as the following functions in $C(M)$:
\begin{equation*}
P=\lambda^{-1}\eta_1\;\;,\qquad\quad E=\lambda^{-1}\log\eta_0\;\;.
\end{equation*}
Again, the presence of $\lambda$ is for dimensional reasons and we choose $E$
as a logarithm because we want it in $\R$, and not in $\R^+$.

The commutators defining phase-space are
\begin{equation}\label{eq:ps}
[x,E]=0\;\;,\qquad [x,P]=i=[t,E]\;\;,\qquad [t,P]=i\lambda P\;\;,
\end{equation}
and reduce to the classical Heisenberg algebra for \mbox{$\lambda=0$}
(physically, this can be interpreted as a ``low-energy limit'',
i.e.~as an approximation for $|\lambda E|,|\lambda P|\ll 1$).

To complete the picture, we want to define the analogue of a boost generator:
the generator of a transformation of $M$ with a fixed point (of course,
we choose the unit element $(\eta_0,\eta_1)=(1,0)$ as fixed point).

If we define
\begin{equation*}
iN=E\partial_P+P\partial_E\equiv\eta_0\eta_1\partial_0+\log\eta_0\partial_1
\quad,
\end{equation*}
$N$ acts on $E$ and $P$ as a classical boost, and $\{E,P,N\}$ generate
the (undeformed) Poincar\'e algebra:
\begin{equation*}
[N,E]=-iP\;\;,\qquad [N,P]=-iE\;\;.
\end{equation*}
Since $N\equiv -Pt-(E-\lambda P^2)x$, the action on coordinates is
\begin{equation*}
[N,x]=i(t-\lambda Px)\;\;,\qquad [N,t]=i(1-\lambda E-\lambda^2P^2)x+i\lambda Pt\;\;.
\end{equation*}
Thus, the action on coordinates is non-linear and reduces to a classical boost
for $\lambda\to 0$.

Like in the Snyder's case, we have deformed $\R^2$ into a noncommutative
space without breaking the Poincar\'e symmetry.

In the Snyder case, the coordinates are vector fields on $SO(2,1)$ ortogonal
to the submanifold $SO(1,1)$, they are a $G$-algebra module for Poincar\'e and
don't close an algebra.

In the $\kappa$-Minkowski case, the Lie group associated to $\{E,P,N\}$ is
Poincar\'e, isomorphic to $SO(1,1)\ltimes Sb(2,\R)$, where $SO(1,1)\!\simeq\!\R$
is the subgroup generated by $N$. Coordinates are vector fields on $M=Sb(2,\R)\simeq
\{SO(1,1)\ltimes Sb(2,\R)\}/SO(1,1)$, and close a Lie algebra isomorphic to
``$i\cdot sb(2,\R)$''. The boosts $SO(1,1)$ are the isotropy
transformations of $M$, that we identify with momentum space. The Poincar\'e
symmetry of spacetime is obtained by dualizing the action of $SO(1,1)$
and taking the cross-product with momenta.

All what done in this subsection can be generalized to $n+1$ dimensions.
Just substitute $M$ with the matrix group of elements
\begin{equation*}
\ma{e^a & b \\ 0 & \openone}\quad\qquad,\;\textrm{with $a\in\R$, $b\in\R^n$, and
$\openone$ the $n\times n$ identity matrix.}
\end{equation*}
The associated Lie algebra is again called $\kappa$-Minkowski~\cite{Maj94-Bcp}
and arises in quantum group theory as a quantum homogeneous space.

\subsection{Quantum group of symmetries for $\kappa$-Minkowski}\noindent
One could object that the action of the boost $N$ on $\kappa$-Minkowski
coordinates depends on momenta. In more mathematical terms: spacetime is not an algebra
module for Poincar\'e, although phase-space is.

It is obvious that a noncommutative space cannot be a quantum homogeneous space
for a Lie group (I mean `embeddable', see appendix~\ref{app}). Or, in other
words, that the algebra of functions on a Lie group has only commutative subgroups.

It results that $\kappa$-Minkowski is embeddable in a quantum group,
$\kappa$-Poincar\'e, whose $\lambda\to 0$ limit is the classical Poincar\'e
group. This fact will be briefly discussed in the present subsection.

We work first in the Euclidean case, and then pass to the pseudo-Euclidean version
through a Wick rotation.

Let $z=x+it$. The algebra of coordinates is generated by $z$ and $z^*$, with relation
\begin{equation}\label{eq:kMin2D}
[z,z^*]=\lambda(z+z^*)\;\;.
\end{equation}
We write the coaction of a yet-unidentified Hopf algebra, generated by
$\{w,w^*,a,a^*\}$, as
\begin{equation*}
\Delta_L(z)=w\otimes z+a\otimes 1\;\;.
\end{equation*}
This extends to a \mbox{$*$-morphism} if it is compatible with the commutator
(\ref{eq:kMin2D}). Using the rule $[A\otimes B,C\otimes D]=[A,C]\otimes BD
+CA\otimes [B,D]$, we write this condition as
\begin{align*}
[\Delta(z),\Delta(z^*)]=[w,w^*]\otimes zz^*+(\lambda w^*w+[w,a^*])\otimes z^*+
(\lambda w^*w+[w,a^*])^*\otimes z\,+ \\ +\,[a^*,a]\otimes 1
\equiv\lambda\bigl(w\otimes z+w^*\otimes z^*+(a+a^*)\otimes 1\bigr)
=\Delta([z,z^*])\;\;.
\end{align*}
The commutator (\ref{eq:kMin2D}) is covariant if and only if
\begin{subequations}\label{eq:ft}
\begin{align}
[w,w^*] &=0 \;\;,       & [a^*,a] & =\lambda(a+a^*) \;\;, \\
[w,a^*] &=\lambda(1-w^*)w \;\;, \hspace{-1cm}
 & [w^*,a] & =-\lambda w^*(1-w)\;\;.
\end{align}
\end{subequations}
Notice that the map $z\mapsto a$ extends to a $*$-algebra morphism between
$\kappa$-Minkowski and a subalgebra of the symmetry Hopf-algebra. This means
that \emph{$\kappa$-Minkowski is an embeddable homogeneous space} for a suitable
Hopf algebra that we are going to construct. The further
condition $w^*w=1$ ensures the existence of the antipode and of a quadratic
invariant (i.e.~$z^*z=x(x+\lambda)+t^2$). The last two conditions in (\ref{eq:ft})
are one the conjugate of the other, and both are equivalent to
the condition $[w,a]=w[w^*w,a]-w[w^*,a]w\equiv -w[w^*,a]w=\lambda(1-w)w$.

To conclude, the Hopf algebra of symmetries is generated by $w$, $a$ and their
conjugates, with relations:
\begin{align*}
w^*w &=ww^*=1 \;\;, \\ [a^*,a] &=\lambda(a+a^*) \;\;, \\ [w,a] &=\lambda(1-w)w \;\;.
\end{align*}
The coproduct, counit and antipode are the standard ones for a matrix
quantum group, given by the well-looking formulas:
\begin{align*}
\Delta\ma{w & a \\ 0 & 1} &=\ma{w & a \\ 0 & 1}\dt\ma{w & a \\ 0 & 1} \;\; , \\
\epsilon\ma{w & a \\ 0 & 1} &=\ma{1 & 0 \\ 0 & 1} \;\; , \\
S\ma{w & a \\ 0 & 1} &=\ma{w & a \\ 0 & 1}^{-1}\equiv\ma{w^* & -w^*a \\ 0 & 1} \;\; ,
\end{align*}
and telling us that $\ma{w & a \\ 0 & 1}$ is a corepresentation (the fundamental one).

This Hopf algebra is the bicrossproduct~\cite{HA} of a
classical rotation subgroup $U(1)$ and $U(sb(2,\R))$. The dual Hopf algebra is
generated by three real elements $\{N,E,P\}$ (for notational convenience we continue
to use the same symbols of sec.~\ref{sec:sym}, but we remark that the Hopf algebra is
different).

The $\{E,P\}$ subalgebra is dual to $\kappa$-Minkowski, and can be determined
(in an equivalent manner) as follows. Since $\kappa$-Minkowski is a classical
Lie-algebra, the dual is the coordinate Hopf algebra of the group of elements
$e^{ip_1 x}e^{ip_0t}$, where $(p_0,p_1)\in\R^2$ and $e:=\exp$ being the
exponential map (this is the connected component $M$ of the group of affine
transformations of $\R$, just the group which we started from). We define $E$
and $P$ to be the following functions:
\begin{equation*}
E(e^{ip_1 x}e^{ip_0 t})=p_0 \;\;,\qquad P(e^{ip_1 x}e^{ip_0t})=p_1 \;\;.
\end{equation*}
Of course, $[E,P]=0$. The product rule between elements of the group (given by
the Baker-Campbell-Hausdorff formula and providing an obvious isomorphism with
$M$):
\begin{equation*}
(e^{ip_1 x}e^{ip_0 t})(e^{ip'_1 x}e^{ip'_0t})=
 e^{i(p_1+p_1'e^{\lambda p_0})x}e^{i(p_0+p_0')t}\;\;,
\end{equation*}
tells us that the coproduct, dual to the product, is determined by the
equations
\begin{align*}
(\Delta E)(e^{ip_1 x}e^{ip_0 t},e^{ip'_1 x}e^{ip'_0t}) &=
  E(e^{i(p_1+p_1'e^{\lambda p_0})x}e^{i(p_0+p_0')t})=p_0+p_0' \;\;, \\
(\Delta P)(e^{ip_1 x}e^{ip_0 t},e^{ip'_1 x}e^{ip'_0t}) &=
  P(e^{i(p_1+p_1'e^{\lambda p_0})x}e^{i(p_0+p_0')t})=p_1+p_1'e^{\lambda p_0} \;\;.
\end{align*}
Then, one has:
\begin{align*}
\Delta E &=\Da{E} \;\;, \\
\Delta P &=P\otimes 1+e^{\lambda E}\otimes P \;\;.
\end{align*}
Similarly, dualizing unit and inverse, one calculate counit and antipode
\begin{equation*}
\epsilon(E)=\epsilon(P)=0 \;\;,\qquad S(E)=-E \;\;,\qquad S(P)=-Pe^{-\lambda E} \;\;.
\end{equation*}
The full Hopf-algebra of symmetries is the bicrossproduct dual to $\C[U(1)]\bcp
U(sb(2,\R))$. Explicitly, the commutators are:
\begin{equation*}
[N,P]=\tfrac{i}{2\lambda}\bigl(1-e^{2\lambda E}\bigr)+\tfrac{i\lambda}{2}P^2
\;\;,\qquad [N,E]=iP\;\;,
\end{equation*}
while $\,\epsilon(N)=0$, $\,S(N)=-e^{-\lambda E}N\,$ and
\begin{equation*}
\Delta(N)=N\otimes 1+e^{\lambda E}\otimes N \;\;.
\end{equation*}
Notice that in the basis
\begin{equation*}
P_x=E \;\;,\qquad P_y=Pe^{-\lambda E/2} \;\;,\qquad J=Ne^{-\lambda E/2}\;\;,
\end{equation*}
the Hopf--algebra is just the \emph{quantum Euclidean group} derived
in~\cite{Cel90-E2} by contraction of $U_q(su(2))$.
The pseudo--Euclidean version is the Hopf algebra generated by $\{N',E',P'\}$
with relations:
\begin{equation*}
[E',P']=0\;\;,\qquad
[N',P']=\tfrac{i}{2\lambda}\bigl(1-e^{2\lambda E'}\bigr)-\tfrac{i\lambda}{2}P'^2
\;\;,\qquad [N',E']=-iP' \;\;,
\end{equation*}
and with the same coproduct as in the Euclidean case. This is the two-dimensional
version of $\kappa$-Poincar\'{e}, obtained for the first time in~\cite{Luk91-qD}
by contraction of $U_q(so(3,2))$.

Dualizing the coaction and passing to the pseudo--Euclidean signature, we find that
the action is standard:
\begin{align*}
N'\az x &=it\;\;,   & P'\az x&=-i \;\;,  & E'\az x&=0  \;\;, \\
N'\az t &=ix\;\;,   & P'\az t&=0  \;\;,  & E'\az t&=-i \;\;.
\end{align*}
The action is undeformed, but the coproduct is. For this reason,
one can calculate
\begin{align*}
N'\az x^2 &=(N'\az x)x+x(N'\az x)=i(tx+xt) \;\;, \\
N'\az t^2 &=(N'\az t)t+(e^{\lambda E'}\az t)(N'\az t)=
         i(tx+xt)+\lambda x=N'\az x^2-i\lambda N'\az t \;\;,
\end{align*}
and prove that the quadratic invariant is deformed into $x^2-t(t+i\lambda)$,
that is:
\begin{equation*}
N'\az \{ x^2-t(t+i\lambda) \} =0 \;\;.
\end{equation*}

Finally, a phase--space can be obtained as a cross--product of momenta and
coordinates. The relations defining the phase-space are
\begin{equation*}
fg=(f_{(1)}\az g)f_{(2)} \;\;,
\end{equation*}
for $f$ a generic function of the momenta, $g$ a function of the coordinates
and $\Delta T=T_{(1)}\otimes T_{(2)}$ the Sweedler notation~\cite{HA}.
Thus:
\begin{equation*}
[x,P']=[t,E']=i \;\;,\qquad [t,P']=i\lambda P' \;\;,\qquad [x,E']=0\;\;.
\end{equation*}
This is the same algebra defined by eq.~(\ref{eq:ps}).

A second (different) phase--space can be constructed via a cross-product
using the right canonical action of momenta on coordinates, instead of the left one.

The construction of phase space as a cross-product (in $3+1$ dimensions)
was perfomed in~\cite{Luk,Now}, an analysis of physical consequences
is in~\cite{GAC}.

\section{Spectral geometry of $\kappa$-Minkowski}\label{sec:cinque}\noindent
Let us quote from~\cite{ConRe} the definition of spectral triple, noncommutative
generalization of the notion of Riemannian spin$^c$ manifold.

\begin{df}
A spectral triple $(\A,\HH,D)$ is given by an involutive algebra $\A$, a
representation $\pi:\A\to\B(\HH)$ by bounded operators on a Hilbert space $\HH$,
and a self-adjoint operator $D=D^*$ with dense domain in $\HH$, such that:
\begin{itemize}
\item[{\bf (S1)}] $\;\;\pi(a)(D^2+1)^{-1/2}$ is a compact operator for all $a\in\A\,$;
\item[{\bf (S2)}] $\;\;[D,\pi(a)]$ is a bounded operator for all $a\in\A\,$.
\end{itemize}
\end{df}

A commutative example is the canonical spectral triple associated to the
spin structure of $\R^2$: $(C^\infty_0(\R^2),\HH,\D)$,
with $C^\infty_0(\R^2)$ smooth functions vanishing at infinity on $\R^2$,
$\HH=L^2(\R^2)\otimes\C^2$ the Hilbert space of $L^2$-spinors and $\D$ the
Dirac operator:
\begin{equation*}
\D=i(\sigma_1\partial_0+\sigma_2\partial_1)
  =\ma{0 & i\partial_0+\partial_1 \\ i\partial_0-\partial_1 & 0} \;\;.
\end{equation*}
Here $\sigma_j$ are the Pauli matrices, $\eta_\mu$ coordinates in $\R^2$
and $\partial_\mu:=\partial/\partial\eta_\mu$ the corresponding derivatives.
Greek letters run over $0,1$. We define also $\vec{\eta}=(\eta_0,\eta_1)$.

$C^\infty_0(\R^2)$ is a Frech\'et pre-$C^*$-algebra. The supremum norm
is equivalent to the operator norm on $\HH$, and the $C^*$-algebra
completion is $C_0(\R^2)$, the algebra of continuous functions vanishing
at infinity.

\begin{df}
A spectral triple $(\A,\HH,D)$ is \emph{even} if exists a grading
$\gamma\in\B(\HH)$, $\gamma=\gamma^*$ and $\gamma^2=1$, such that
$\gamma D=-D\gamma\,$ and $\,a\gamma=\gamma a\;\forall\;a\in\A$.
\end{df}

The operator:
\begin{equation}\label{eq:gamma}
\gamma:=\ma{1 & 0 \\ 0 & -1\,}\;\;,
\end{equation}
is a grading for the canonical spectral triple on $\R^2$.

We quote also the notion of equivariance with respect to the action
of a Lie group.

\begin{df}
Let $H$ be Lie group, $\rho:H\to\B(\HH)$ a representation and $\;\az:H\times\A\to\A$
a covariant action. An even spectral triple $(\A,\HH,D,\gamma)$ is $H$-equivariant if:
\begin{itemize}
\item[{\bf (E1)}] $\;\;\rho(h)\pi(a)\rho(h)^{-1}=\pi(h\az a)\;$ for all
 $\;a\in\A\,$, $h\in H\,$;
\item[{\bf (E2)}] $\;\;\rho(h)D\rho(h)^{-1}=D\;$ and $\;\rho(h)\gamma\rho(h)^{-1}=\gamma\;$
 for all $\;h\in H\,$.
\end{itemize}
\end{df}

The spin structure of $\R^2$ is equivariant with respect to the spin representation
of $H:=ISO(2)\simeq SO(2)\ltimes\R^2$, the group of isometries of the Euclidean plane $\R^2$.

If $(R,\vec{v}),(R',\vec{v}\,')\in SO(2)\times\R^2$,
the multiplication law of $ISO(2)$ is:
\begin{equation*}
(R,\vec{v})\cdot (R',\vec{v}\,')=(RR',\vec{v}+R\vec{v}\,')\;\;,
\end{equation*}
and the spin representation $\rho:H\to\B(\HH)$ is defined by:
\begin{equation}\label{eq:spin}
\{\rho(R,\vec{v})\psi\}(\vec{\eta}):=
R\cdot\psi\bigl(R^{-1}(\vec{\eta}-\vec{v})\bigr)
\;\;,\qquad\forall\; \psi\in\HH\;\;.
\end{equation}

The representation $\pi$ of the algebra satisfies (E1) if we
define $\az$ to be the pull-back of the natural action on $\R^3$:
\begin{equation*}
\{(R,\vec{v})\az f\}(\vec{\eta})=f\bigl(R^{-1}(\vec{\eta}-\vec{v})\bigr)
\;\;,\qquad\forall\; f\in C^\infty_0(\R^2)\;\;.
\end{equation*}

Differentiating the action of the group, we arrive at the equivalent
notion of $U(\mathrm{Lie}H)$-equivariance. This notion can be generalized
to a generic Hopf-algebra.

\begin{df}
Let $\U$ be an Hopf-algebra, $\rho:\U\to\B(\HH)$ a representation and
$\;\az:\U\times\A\to\A$ a covariant action. An even spectral triple
$(\A,\HH,D,\gamma)$ is $\U$-equivariant if:
\begin{itemize}
\item[{\bf (E1$^\prime$)}] $\;\;\rho(u_{(1)})\pi(a)\rho(Su_{(2)})=\pi(u\az a)\;$ for all
 $\;a\in\A\,$, $u\in\U\,$;
\item[{\bf (E2$^\prime$)}] $\;\;\rho(u)D=D\rho(u)\;$ and $\;\rho(u)\gamma=\gamma\rho(u)\;$
 for all $\;u\in\U\,$.
\end{itemize}
\end{df}

In the following we construct the algebras replacing continuous
and smooth functions vanishing at infinity associated to $\kappa$-Minkowski,
and describe the spinor representation.

Since the metric properties of the space depend on the Dirac
operator, at this point there is no difference between Euclidean
and Lorenzian version.

We then analyze the problem of constructing an Euclidean Dirac
operator. We exibith an operator that is essentially the
Dirac operator on the commutative subspace $\R$ of $\kappa$-Minkowski,
and prove that it defines a spectral triple. We search also for an
equivariant Dirac operator and find that there is just one,
and does not satisfy the axioms of a spectral triple.

\subsection{The algebra of polynomials of the ``noncommutative coordinates''}\noindent
Let us recall the construction of the algebra of coordinates on
$\kappa$-Minkowski.

We call $G:=(\R^2,\dotplus)$ the space $\R^2$ with deformed sum:
\begin{equation*}
(p_0,p_1)\dotplus (p'_0,p'_1)=(p_0+p'_0,p_1+p'_1e^{\lambda p_0})\qquad
\forall\;(p_0,p_1),(p'_0,p'_1)\in\R^2\;.
\end{equation*}
The map:
\begin{equation*}
(p_0,p_1)\mapsto g:=\ma{e^{\lambda p_0} & p_1 \\ 0 & 1}\;\;,
\end{equation*}
is an isomorphism between $G$ and $\mathrm{Aff}_0(\R)$, (the connected component of)
the group of affine transformation of $\R$. We will identify $G$ and $\mathrm{Aff}_0(\R)$.

There is only one unitary irreducible infinite-dimensional representation
of the group $\mathrm{Aff}(\R)$, defined on the Hilbert space
$L^2(\R^*,\de\eta_1/|\eta_1|)$ by~\cite{Gel}:
\begin{equation*}
\{g\cdot\varphi\}(\eta_1):=e^{ib\eta_1}\varphi(a\eta_1)\qquad
,\;\varphi\in L^2(\R^*,\de\eta_1/|\eta_1|)\;,\;g=\ma{a & b \\ 0 & 1}\in
\;\mathrm{Aff}(\R).
\end{equation*}
We take the tensor product of this representation with the trivial representation
$g\mapsto e^{i\eta_0p_0}$ of the abelian subgroup $\R$. Then we take the direct
integral on $\eta_0\in\R$ with an arbitrary measure $\de\mu(\eta_0)$ on $\R$.

The result is a unitary representation of $G$ on the space:
\begin{equation}\label{eq:H}
\HH_\mu:=L^2(\R\times\R^*,\de\mu(\eta_0)\,|\eta_1|^{-1}\de\eta_1)\otimes\C^2\;\;,
\end{equation}
defined by:
\begin{align}\label{eq:pi}
\pi: &\; G\to\B(\HH_\mu)\;\;,\quad\vec{p}\mapsto\pi(\vec{p}\,) \;\;, \nonumber \\
\rule{0pt}{3ex} & \{\pi(\vec{p}\,)\psi\}(\vec{\eta})=e^{i\vec{p}\cdot\vec{\eta}}
        \psi(\eta_0,\eta_1e^{\lambda p_0})\;\;,
\end{align}
where $\psi\in\HH_\mu$. One can explicitly check that it is an homomorphism:
\begin{equation*}
\pi(\vec{p}\,)\pi(\vec{p}\,')=\pi(p_0+p_0',p_1+p'_1e^{\lambda p_0})\;\;.
\end{equation*}
We indicate with $\HH_0$ the Hilbert space obtained taking as
$\de\mu(\eta_0)$ the Lebesgue measure $\de\eta_0$.

In the previous section, we have defined the ``noncommutative coordinates''
on $\kappa$-Minkowski as the vector fields associated to the generators of $U(\g)$:
\begin{equation*}
\ma{\lambda & 0 \\ 0 & 0}\;\;,\qquad\ma{0 & 1 \\ 0 & 0}\;\;,
\end{equation*}
where $\g$ is the Lie algebra of $G$.

We can obtain a representation of these coordinates on $\HH_\mu$ using
the differential of $\pi$:
\begin{align*}
\hat{x}_0 &=i\,\de\pi\ma{\lambda & 0 \\ 0 & 0}=\eta_0-i\lambda
            \eta_1\partial_1   \;\;, \\
\hat{x}_1 &=i\,\de\pi\ma{0 & 1 \\ 0 & 0}=\eta_1   \;\;.
\end{align*}
The operators $(\hat{x}_0,\hat{x}_1)$ generate an algebra isomorphic
to the $\kappa$-Minkowski algebra:
\begin{equation}\label{eq:cr}
[\hat{x}_0,\hat{x}_1]=-i\lambda\hat{x}_1  \;\;.
\end{equation}

We indicate with $\R^2_\lambda$ the virtual ``quantum space''
associated to the algebra $U(\g)$, and call elements of $U(\g)$
the `polynomial functions' on $\kappa$-Minkowski space.

When $\lambda=0$, the representation $\de\pi$ reduces to the ordinary
(unbounded) representation of polynomial functions on $\R^2$
via pointwise multiplication on $\HH_\mu$.

\subsection{Continuous and smooth `functions' on $\kappa$-Minkowski space}\noindent
To construct a spectral triple with the polynomial algebra $U(\g)$
is problematic, since $\hat{x}_\mu$ cannot be represented
by bounded operators, and $(\ref{eq:cr})$ can be satisfied only on a
dense domain in $\HH_\mu$. It is the same problem one encounters in the
canonical quantization of phase-space in quantum-mechanics.
A possible solution is to shift the attention from $\hat{x}_\mu$
to complex exponentials, that is, from the Lie algebra $\g$ to
the Lie group $G$. This is Weyl quantization, defined as a map
associating complex exponentials on $\R^2$ to elements of $G$,
represented by unitary operators on $\HH_\mu$. The quantization map
can be extended to an involutive subalgebra of $C_0(\R^2)$ using
Fourier transform.

We call $\Fn$ the following class of functions:
\begin{equation*}
\Fn:=C_0(\R^2)\cap\HH_0\;\;,
\end{equation*}
and define the following quantization map:
\begin{align*}
\Omega:&\;\Fn\to\B(\HH_\mu)\;\;, \\
\rule{0pt}{3ex} & f\mapsto\Omega(f)=\int_{\R^2}\de^2p\,\tilde{f}(\vec{p}\,)\pi(\vec{p}\,)\;\;,
\end{align*}
where $\pi$ is the representation (\ref{eq:pi}) and $\tilde{f}$ is the
Fourier transform of $f$:
\begin{equation*}
\tilde{f}(\vec{p}\,)=\frac{1}{(2\pi)^2}\int_{\R^2}
f(\vec{\eta}\,)e^{-i\vec{p}\vec{\eta}}\de^2\eta\;\;.
\end{equation*}

Although $f=e^{i\vec{p}\vec{\eta}}\notin C_0(\R^2)$, one can formally
verify taking $\tilde{f}(\vec{p}\,')=\delta^{(2)}(\vec{p}-\vec{p}\,')$ that:
\begin{equation*}
\Omega(e^{i\vec{p}\vec{\eta}})=\pi(\vec{p}\,) \;\;,
\end{equation*}
and since $\eta_\mu=-i\frac{\partial}{\partial p_\mu}\big|_{\vec{p}=0}e^{i\vec{p}\vec{\eta}}$,
\begin{equation*}
\Omega(\eta_\mu)=-i\tfrac{\partial}{\partial p_\mu}\big|_{\vec{p}=0}\pi(\vec{p}\,)
\equiv\hat{x}_\mu \;\;.
\end{equation*}

The explicit expression of the quantization map is:
\begin{equation}\label{eq:rep}
\{\Omega(f)\psi\}(\vec{\eta})=\int_{\R^2}\tilde{f}(\vec{p}\,)e^{i\vec{p}\vec{\eta}}
\psi(\eta_0,\eta_1e^{\lambda p_0})\de^2 p\qquad,\;\;f\in\Fn\;,\;\;\psi\in\HH_\mu\;,
\end{equation}
and we need to verify that it defines bounded operator.

\begin{prop}
The quantization map $\Omega$, defined by (\ref{eq:rep}), sends $\Fn$ into bounded
operators $\B(\HH_\mu)$. For this class of functions, the operator norm is bounded by:
\begin{equation}\label{eq:upper}
||\Omega(f)||^2 :=\sup_{\psi\in\HH_\mu,\psi\neq 0}
\frac{||\Omega(f)\psi||^2_{\HH_\mu}}{||\psi||^2_{\HH_\mu}}
\leq\frac{1}{2\pi}\,||f||_{\HH_0}^2\;\;,
\end{equation}
where $||_{\HH_\mu}$ indicate the norm in the Hilbert space $\HH_\mu$.
\end{prop}

\proof{Let $a=e^{\lambda p_0}$ and call:
\begin{equation*}
F(\vec{\eta},a)=\frac{1}{2\pi}\int_{\R}\de\eta'_0\,f(\eta'_0,\eta_1)e^{ip_0(\eta_0-\eta'_0)}
   =\int_{\R}\de\eta_1\,\tilde{f}(\vec{p})e^{i\vec{p}\vec{\eta}}   \;\;.
\end{equation*}
Then:
\begin{equation*}
\{\Omega(f)\psi\}(\vec{\eta})=\int_{\R}\de p_0\,F(\vec{\eta},a)\psi(\eta_0,a\eta_1)
=\int_{\R^+}\frac{\de a}{\lambda a}\,F(\vec{\eta},a)\psi(\eta_0,a\eta_1)\;\;.
\end{equation*}
Since (partial) Fourier transform is an isometry of $L^2$, then
$F\in L^2(\R,\de p_0)=L^2(\R^+,\de a/\lambda a)$ for each fixed
$\vec{\eta}$. Using Schwartz inequality:
\begin{align}
|\Omega(f)\psi(\vec{\eta})|^2 &\leq\left(\int_{\R^+}\frac{\de a}{\lambda a}\,
   \big|F(\vec{\eta},a)\psi(\eta_0,a\eta_1)\big|\right)^2 \nonumber \\
&\leq\left(\int_{\R^+}\frac{\de a}{\lambda a}\,|F(\vec{\eta},a)|^2\right)
   \left(\int_{\R^+}\frac{\de a}{a}\,|\psi(\eta_0,a\eta_1)|^2\right) \;\;.
  \label{eq:ineqA}
\end{align}
Now:
\begin{subequations}\label{eq:ineqB}
\begin{align}
\int_{\R^+}\frac{\de a}{\lambda a}\,|F(\vec{\eta},a)|^2
&=\frac{1}{(2\pi)^2}\int_{\R^+}\frac{\de a}{\lambda a}\,
  \left|\int_{\R}\de\eta'_0\,f(\eta'_0,\eta_1)e^{ip_0(\eta_0-\eta'_0)}\right|^2 \nonumber \\
&=\frac{1}{(2\pi)^2}\int_{\R}\de p_0\int_{\R^2}\de\eta'_0\de\eta''_0\,\,
  \overline{f}(\eta'_0,\eta_1)f(\eta''_0,\eta_1)e^{ip_0(\eta'_0-\eta''_0)} \nonumber \\
&=\frac{1}{2\pi}\int_{\R}\de\eta'_0\,|f(\eta'_0,\eta_1)|^2 \;\;, \\
\int_{\R^+}\frac{\de a}{a}\,|\psi(\eta_0,a\eta_1)|^2 &\leq
\int_{\R^*}\frac{\de a}{|a|}\,|\psi(\eta_0,a\eta_1)|^2=
\int_{\R^*}\frac{\de\eta'_1}{|\eta'_1|}\,|\psi(\eta_0,\eta'_1)|^2 \;\;,
\end{align}
\end{subequations}
where $\eta'_1=a\eta_1$ and in the last step we used dilatation
invariance of the measure.

Using the inequalities (\ref{eq:ineqA}) and (\ref{eq:ineqB}) we arrive at:
\begin{align*}
||\Omega(f)\psi||^2_{\HH_\mu} &=
\int_{\R\times\R^*}\frac{\de\mu(\eta_0)\,\de\eta_1}{|\eta_1|}\,|\Omega(f)\psi(\vec{\eta})|^2 \\
&\leq\frac{1}{2\pi}\left(\int_{\R\times\R^*}\frac{\de\eta'_0\de\eta_1}{|\eta_1|}\,
|f(\eta'_0,\eta_1)|^2\right)\left(\int_{\R\times\R^*}\frac{\de\mu(\eta_0)\,\de\eta'_1}
{|\eta'_1|}|\psi(\eta_0,\eta'_1)|^2\right) \\ &=\frac{1}{2\pi}\,||f||_{\HH_0}^2\cdot
||\psi||^2_{\HH_\mu}\;\;.
\end{align*}
Using the last inequality, we find the upper bound (\ref{eq:upper}) for the operator norm.}

We define the following $*$-product on $\Fn$:
\begin{equation}\label{eq:star}
f_1*f_2:=\Omega(f_1)f_2\;\;,\qquad\forall\;f_{1,2}\in\Fn\; .
\end{equation}
Let $\Fns$ be the subspace of smooth functions of $\Fn$.

\begin{prop}\label{prop:A}
$\,\A:=(\Fn,*)$ and $\A^\infty:=(\Fns,*)\subset\A$ are involutive
algebras (associative without unit). $\Omega:\A\to\B(\HH_\mu)$
is a unitary representation.
\end{prop}

\pagebreak

\proof{Since $\Omega(f_1)$ is a bounded operator on $\HH_\mu$,
for all $\mu$, and $f_2\in\HH_0$, the product $f_1*f_2$ is
in $\HH_0$. Moreover, $f_1*f_2\in C_0(\R^2)$ and is smooth
if $f_1$ and $f_2$ are.

By construction:
\begin{equation*}
\Omega(f_1*f_2)=\Omega(f_1)\Omega(f_2)\;\;.
\end{equation*}
This guarantees associativity of the product, and proves
that $\Omega$ is a representation.

Finally, $\Omega(\bar{f}_1)=\Omega(f_1)^+$ is the adjoint of $\Omega(f_1)$,
since the representation $\pi$ of the group $\mathrm{Aff}_0(\R)$
is unitary. So we have a unitary representation of $\A$. }

\bigskip

We call $\A$ the algebra of `continuous functions' on $\kappa$-Minkowski
space, and $\A^\infty$ the subalgebra of `smooth functions'.
Since $1\notin\A$, the space is not compact.

From (\ref{eq:rep}), using Leibniz rule and
the property $ip_\mu\tilde{f}=\widetilde{\partial_\mu f}$, we deduce
the following `deformed' Leibniz rule:
\begin{subequations}\label{eq:Leib}
\begin{align}
\partial_0\big\{\Omega(f_1)f_2\big\} &=\Omega(\partial_0f_1)f_2+
      \Omega(f_1)\partial_0f_2 \;\;, \\
\partial_1\big\{\Omega(f_1)f_2\big\} &=\Omega(\partial_1f_1)f_2+
      \Omega(e^{-i\lambda\partial_0}f_1)\partial_1f_2 \;\;,
\end{align}
\end{subequations}
or equivalently:
\begin{align*}
\partial_0(f_1*f_2) &=(\partial_0f_1)*f_2+f_1*(\partial_0f_2) \;\;, \\
\partial_1(f_1*f_2) &=(\partial_1f_1)*f_2+
      (e^{-i\lambda\partial_0}f_1)*(\partial_1f_2) \;\;,
\end{align*}
for all $f_1,f_2\in\Fns$.

\subsection{A Dirac operator for $\kappa$-Minkowski space}\label{sec:Dirac}\noindent
The first attempt to define a Dirac operator would be to use the
classical one $\D$. \\
Properties (\ref{eq:Leib}) means that $\D$ has not bounded commutator with
the algebra. Indeed:
\begin{equation*}
[\partial_0,\Omega(f)]=\Omega(\partial_0f)\;\;,\qquad f\in\A^\infty\;,
\end{equation*}
is bounded, but $[\partial_1,\Omega(f)]$ is not, due to the presence
of the unbounded operator $e^{-i\lambda\partial_0}$.

$D:=i\partial_0$ has non-trivial sign (it is not positive), has dense
domain in $\HH$ and bounded commutators with $\A$. Geometrically,
the evaluation at $\eta_1=0$:
\begin{equation*}
\Omega(f)\mapsto f(\eta_0,0)\;\;,
\end{equation*}
is an algebra morphism $\A\to C_0(\R)$, and tells us
that $\R$ is a commutative subspace. $D$ is just the Dirac operator
on this subspace.

Let $\Delta:=-\partial_0^2+1$.
Intuitively, one would expect that the axiom (S1) in the definition
of spectral triple is not satisfied. For $\lambda=0$, $f\cdot\Delta^{-1/2}$
is not a compact operator on $L^2(\R^2)$. Surprisingly if $\lambda\neq 0$,
$\Omega(f)\Delta^{-1/2}$ is a compact operator on $\HH_\mu$, if $\mu$
is a finite measure on $\R$ (i.e.~$\int_{\R}\de\mu(\eta_0)<\infty$) and
absolutely continous with respect to the Lebesgue measure.

\begin{prop}
Let $\mu$ be a finite measure on $\R$, absolutely continuous with
respect to the Lebesgue measure, and let $\lambda\neq 0$. Then,
$(\A^\infty,\HH_\mu,i\partial_0)$ is a $1^+$-summable spectral triple.
The associated Dixmier trace is just the cyclic integral of~\cite{DJMTWW}:
\begin{equation}\label{eq:ci}
\nint\Omega(f)\Delta^{-1/2}:=\mathrm{Res}_{z=1}\tr_{\HH_\mu}\Omega(f)\Delta^{-z/2}
\propto\int_{\R\times\R^*}f(\vec{x})\,\frac{\de x_0\de x_1}{|x_1|}\;\;,
\end{equation}
where $\Delta:=-\partial_0^2+1$.
\end{prop}

\proof{We decompose $\HH_\mu=\V_0\otimes\V_1$, with $\V_0:=L^2(\R,\de\mu(\eta_0))$
and $\V_1:=L^2(\R^*,\de\eta_1/|\eta_1|)$.

Let us recall some fact from~\cite{CM} (see also~\cite{Hig} for
a pedagogical presentation and~\cite{Gay03} for considerations on the
non-compact case).

Let $\mu'(\eta_0)=\de\mu(\eta_0)/\de\eta_0$. By hypothesis $\mu'\in C_0(\R^2)$.

Being $i\partial_0$ the Dirac operator on $\R$, $f\Delta^{-1/2}$ is compact
on $L^2(\R,\de\eta_0)$ for all $f\in C_0(\R^2)$. In particular, taking
$f=\mu'$, we prove that $\Delta^{-1/2}$ is compact on $\V_0$.

From the formula:
\begin{equation*}
\mathrm{Res}_{z=1}\tr_{L^2(\R,\de\eta_0)}f\Delta^{-z/2}\propto\int_{\R}f(\eta_0)\de\eta_0\;\;,
\end{equation*}
(we don't care about the proportionality constant, which is independent on $f$) we deduce
\begin{equation*}
\mathrm{Res}_{z=1}\tr_{\V_0}\Delta^{-z/2}=
\mathrm{Res}_{z=1}\tr_{L^2(\R,\de\eta_0)}\mu'\Delta^{-z/2}
\propto\int_{\R}\de\mu(\eta_0)<\infty\;\;.
\end{equation*}
The operator $\Delta^{-z/2}$ is traceclass on $\V_0$ if $z>1$, and in the Dixmier
class $\mathcal{L}^{1+}(\V_0)$ if $z=1$.

On $\HH_\mu$, if $f\in\A$, the kernel of the operator $\Omega(f)\Delta^{-z/2}$
is the distribution:
\begin{equation*}
K_f(\vec{\eta},\vec{\eta}\,')=\frac{1}{2\pi}\int_{\R^2}\de^2p\,\tilde{f}
     (\vec{p})e^{i\vec{p}\vec{\eta}}\delta(\eta'_1-\eta_1e^{\lambda p_0})
     \int_{\R}\de\xi\,(1+\xi^2)^{-z/2}e^{i\xi(\eta_0-\eta'_0)}  \;\;,
\end{equation*}
where the integral in $\de\xi$ is the resolvent of $\Delta^{z/2}$.

We consider first the case in which $f$ is integrable
($f\in L^1(\R^*,\de\eta_0\de\eta_1/|\eta_1|)$).

The partial trace of $\Omega(f)\Delta^{-z/2}$ on $\V_1$ is the operator with kernel:
\begin{align*}
\hat{K}_f(\vec{\eta},\vec{\eta}\,') &=\int_{\R^*}\frac{\de\eta_1}{|\eta_1|}
   K_f(\eta_0,\eta_1;\eta'_0,\eta_1) \\
&=\frac{\,\lambda^{-1}}{2\pi}\int_{\R^*}\frac{\de\eta_1}{|\eta_1|}
     \int_{\R}\de p_1\,\tilde{f}(0,p_1)e^{ip_1\eta_1}
     \int_{\R}\de\xi\,(1+\xi^2)^{-z/2}e^{i\xi(\eta_0-\eta'_0)} \\
&=\frac{\,\lambda^{-1}}{2\pi}\left(\int_{\R\times\R^*}\frac{\de x_0\de x_1}{|x_1|}
     f(\vec{x})\right)\int_{\R}\de\xi\,(1+\xi^2)^{-z/2}e^{i\xi(\eta_0-\eta'_0)}  \;\;.
\end{align*}

Then, as operators on $\V_0$:
\begin{equation}\label{eq:finale}
\tr_{\V_1}\{\Omega(f)\Delta^{-z/2}\}=\frac{\,\lambda^{-1}}{2\pi}
\left(\int_{\R\times\R^*}f(\vec{x})\frac{\de x_0\de x_1}{|x_1|}\right)\Delta^{-z/2}  \;\;.
\end{equation}

From what said above about $\Delta^{-z/2}$, we see that $\Omega(f)\Delta^{-z/2}$
is traceclass on $\HH_\mu$ if $z>1$, and in the Dixmier class $\mathcal{L}^{1+}(\HH_\mu)$
if $z=1$.

Taking the trace on $\V_0$ of (\ref{eq:finale}) and then the residue in $z=1$
we prove (\ref{eq:ci}).

Now, since $\Fn\subset L^2$, integrable functions are dense in the algebra
$\A$. So, $\Omega(f)\Delta^{-1/2}$ is in the closure of $\mathcal{L}^{1+}(\HH_\mu)$
for all $f\in\A$. The closure of the Dixmier class are the compact operators $\K$, and
this concludes the proof. }

\medskip

\noindent\textbf{Remark:} The spectral triple constructed in this section has not
a commutative analogue. Axiom (S1) is not satisfied for $\lambda=0$.

\subsection{Equivariance properties of the representation}\noindent
Let $\A$ be the involutive algebra defined in proposition~\ref{prop:A},
$\HH$ the Hilbert space:
\begin{equation*}
\HH:=\HH_0\otimes\C^2=L^2(\R\times\R^*,|\eta_1|^{-1}\de\eta_0\de\eta_1)\otimes\C^2\;\;,
\end{equation*}
and $\gamma$ the grading in (\ref{eq:gamma}). We lift trivially the
representation (\ref{eq:rep}) of $\A$ from $\HH_0$ to $\HH$.

We postpone the problem of constructing an equivariant Dirac operator and
study the equivariance properties of the data $(\A,\HH,\gamma)$.
Clearly $\A$ commutes with $\gamma$, and so $\gamma$ is a natural
candidate for the grading.

The space $\HH_0$ carries a representation of the quantum Euclidean group,
the Euclidean analogue of $\kappa$-Poincar\'e, which we indicate with
$U_\kappa(iso(2))$. Recall that the Hopf algebra is generated by $E,P,N$
with commutation relations:
\begin{equation*}
[E,P]=0\;\;,\qquad [N,P]=\tfrac{i}{2\lambda}\bigl(1-e^{2\lambda E}\bigr)
+\tfrac{i\lambda}{2}P^2\;\;,\qquad [N,E]=iP\;\;,
\end{equation*}
coproduct:
\begin{align*}
\Delta E &=\Da{E} \;\;, \\
\Delta P &=P\otimes 1+e^{\lambda E}\otimes P \;\;, \\
\Delta N &=N\otimes 1+e^{\lambda E}\otimes N \;\;,
\end{align*}
and counit/antipode:
\begin{equation*}
\epsilon(E)=\epsilon(P)=\epsilon(N)=0 \;\;,\quad S(E)=-E \;\;,\quad
S(P)=-Pe^{-\lambda E} \;\;,\quad S(N)=-Ne^{-\lambda E} \;\;.
\end{equation*}

There is a natural representation of the $(E,P)$ sub-Hopf-algebra
of $U_\kappa(iso(2))$ on a space dense in $\HH_0$, defined by
\begin{equation*}
\rho(E)=-i\partial_0\;\;,\qquad\rho(P)=-i\partial_1\;\;.
\end{equation*}
This extend to a representation of the full Hopf-algebra if we
define:
\begin{align}
\rho(N) &=\eta_0\,\rho(P)+\eta_1\,\rho\!\left(\tfrac{1-e^{2\lambda E}}{2\lambda}+
  \tfrac{\lambda}{2}P^2\right) \nonumber \\
&=-i\eta_0\partial_1+\tfrac{1}{2\lambda}\eta_1
  (1-e^{-2i\lambda\partial_0})-\tfrac{\lambda}{2}\eta_1\partial_1^2\;\;. \label{eq:Nrep}
\end{align}
On the space $\C^2$, we call $\sigma$ the representation of the commutative
$\R$ subalgebra, defined by:
\begin{equation*}
\sigma(N):=\tfrac{1}{2}\gamma=\frac{1}{2}\ma{1 & 0 \\ 0 & -1\,}\;\;,
\qquad\sigma(E)=\sigma(P)=0\;\;.
\end{equation*}
On the space $\HH=\HH_0\otimes\C^2$ we consider the representation
$\rho\otimes\sigma$ defined through the (opposite) coproduct:
\begin{equation*}
(\rho\otimes\sigma)(h)=\rho(h_{(2)})\otimes\sigma(h_{(1)})\;\;,\qquad
\forall\;h\in U_\kappa(iso(2))\;.
\end{equation*}
This representation commutes with the grading $\gamma$, since the image
through $\sigma$ of the algebra is in the subspace of $\mathrm{Mat}_2(\C)$
spanned by $1$ and $\gamma$.

Before discussing the $U_\kappa(iso(2))$-equivariance of $(\A,\HH,\gamma)$
we need to define a covariant action of the Hopf algebra on $\A$. If we define:
\begin{equation*}
h\az\Omega(f):=\Omega\bigl(\rho(h)f\bigr)\;\;,\qquad\forall\;h\in U_\kappa(iso(2)),\;
\Omega(f)\in\A\;,
\end{equation*}
this is a representation of the Hopf-algebra. We want to prove that it is
covariant, that is:
\begin{equation*}
h\az\Omega(f_1)\Omega(f_2)=\big\{h_{(1)}\az\Omega(f_1)\big\}\big\{h_{(2)}\az\Omega(f_2)\big\}\;\;,
\end{equation*}
or equivalently using the $*$-product:
\begin{equation*}
\rho(h)(f_1*f_2)=\big\{\rho(h_{(1)})f_1\big\}*\big\{\rho(h_{(2)})f_2\big\}\;\;.
\end{equation*}

\begin{prop}
The action $\az$ of $U_\kappa(iso(2))$ on $\A$ is covariant. Moreover:
\begin{equation*}
h\az a=\rho(h_{(1)})\,a\,\rho(Sh_{(2)})\;\;,
\end{equation*}
for all $h\in U_\kappa(iso(2))$ and $a\in\A$ (on a subspace dense in $\HH_0$).
\end{prop}
\proof{Using (\ref{eq:star}), we deduce that the covariance condition is equivalent to:
\begin{equation}\label{eq:cov}
\rho(h)\Omega(f_1)f_2=\Omega\big(\rho(h_{(1)})f_1\big)\cdot\big\{\rho(h_{(2)})f_2\big\}\;\;.
\end{equation}
Since on both sides appear representations of $U_\kappa(iso(2))$,
it is sufficient to do the check for the generators $E,P,N$.

We can rewrite (\ref{eq:Leib}) as:
\begin{align*}
\rho(E)\Omega(f_1)f_2&=\Omega\big(\rho(E)f_1\big)f_2+
  \Omega(f_1)\big\{\rho(E)f_2\big\}\;\;, \\
\rho(P)\Omega(f_1)f_2&=\Omega\big(\rho(P)f_1\big)f_2+
  \Omega\big(\rho(e^{\lambda E})f_1\big)\big\{\rho(P)f_2\big\}\;\;,
\end{align*}
and then the action of $E$ and $P$ is covariant.

In the same way, using (\ref{eq:Leib}) and (\ref{eq:Nrep}),
it is a straightforward computation to prove the covariance
of the action of $N$.

If in equation (\ref{eq:cov}) we replace $h$ with $h_{(1)}$,
call $a=\Omega(f_1)\in\A$, $f_2=\rho(Sh_{(2)})\psi$, and
recall that $\Omega\big(\rho(h_{(1)})f_1\big)=h_{(1)}\az a$,
we obtain:
\begin{equation*}
\rho(h_{(1)})a\rho(Sh_{(2)})\psi=(h_{(1)}\az a)\rho\big(h_{(2)}S(h_{(3)})\big)
\psi=\big\{\epsilon(h_{(2)})h_{(1)}\az a\big\}\psi=(h\az a)\psi\;\;.
\end{equation*}
This concludes the proof. }

\medskip

Since the representation of $\A$ is lifted diagonally from $\HH_0$
to $\HH=\HH_0\otimes\C^2$, $\sigma(h)a=a\sigma(h)$ for all $h\in U_\kappa(iso(2))$
and $a\in\A$. Then, as a corollary:
\begin{equation*}
h\az a=(\rho\otimes\sigma)(h_{(1)})\,a\,(\rho\otimes\sigma)(Sh_{(2)})\;\;.
\end{equation*}
This means that:
\begin{cor}
$(\A,\HH,\gamma)$ is $U_\kappa(iso(2))$-equivariant.
\end{cor}

\subsection{An $U_\kappa(iso(2))$-equivariant Dirac operator}\noindent
To have an equivariant spectral triple on $\kappa$-Minkowski space, it
remain to find a Dirac operator $D$ that is equivariant for the action of the
quantum Euclidean group. \\
We write $D$ as a formal pseudo-differential operator:
\begin{equation*}
D\psi(\vec{\eta})=\int_{\R^2}e^{i\vec{p}\vec{\eta}}
\ma{0 & T(\vec{\eta},\vec{p}) \\ T(\vec{\eta},\vec{p})^* & 0}
\tilde{\psi}(\vec{p})\de^2p\;\;,\qquad\psi\in\HH\;,
\end{equation*}
and determine the symbol $T$ imposing equivariance. The matrix
form of the symbol is a consequence of the grading and formal
selfadjointness.

Since $(\rho\otimes\sigma)(E)=\rho(E)$,
$(\rho\otimes\sigma)(P)=\rho(P)$ and $(\rho\otimes\sigma)(N)=
\sigma(N)+\rho(N)\sigma(e^{\lambda E})=\rho(N)+\frac{1}{2}\gamma$,
the equivariance conditions become:
\begin{equation*}
[\rho(E),D]=0\;\;,\qquad [\rho(P),D]=0\;\;,\qquad
[\rho(N),D]=-\tfrac{1}{2}[\gamma,D]\equiv -\gamma D\;\;.
\end{equation*}

The first two conditions are equivalent to:
\begin{equation*}
\partial_0 T(\vec{\eta},\vec{p})=\partial_1 T(\vec{\eta},\vec{p})=0 \;\;.
\end{equation*}
Then, $T(\vec{\eta},\vec{p})=T(\vec{p})$. The last condition can be written as:
\begin{equation}\label{eq:last}
[\chi(N),T(\vec{p})]=-T(\vec{p})\;\;,\qquad [\chi(N),T(\vec{p})^*]=T(\vec{p})^*\;\;,
\end{equation}
where
\begin{equation*}
\chi(N)=ip_1\frac{\partial}{\partial p_0}+i\left(\frac{1-e^{2\lambda p_0}}{2\lambda}+
\frac{\lambda}{2}p_1^2\right)\frac{\partial}{\partial p_1} \;\;.
\end{equation*}
If we define $\chi(E)=p_0$ and $\chi(P)=p_1$, then the map $\chi$ extends
to a representation of the $U_\kappa(iso(2))$ Hopf algebra.

In~\cite{Base} was constructed for the first time a (formal) isomorphism
between $\kappa$-Poincar\'e and the Poincar\'e algebra. The Euclidean
conterpart is the change of coordinates:
\begin{equation*}
e^{-\lambda p_0}p_1=:r\sin\theta\;\;,\qquad\tfrac{1}{\lambda}\sinh(\lambda p_0)
+\tfrac{\lambda}{2}p_1^2e^{-\lambda p_0}=:r\cos\theta\;\;,
\end{equation*}
with $r\in\R^+_0$ and $\theta\in S^1$. The Casimir of the quantum Euclidean
group is:
\begin{equation*}
m^2:=\left(\tfrac{2}{\lambda}\sinh\tfrac{\lambda E}{2}\right)^2+e^{-\lambda E}P^2\;\;.
\end{equation*}
It is related to $r^2$ by:
\begin{equation}\label{eq:Cas}
r^2=\chi\Big(m^2\big(1+\tfrac{\lambda^2 m^2}{4}\big)\Big)\;\;,
\end{equation}
and then $r^2$ is central and $[\chi(N),r^2]=0$.
So, $\chi(N)=\tilde{N}(r,\theta)\partial_\theta$ is a
derivation in the $\theta$ direction, with $\tilde{N}$ defined by
$\,\tilde{N}(r,\theta)re^{i\theta}=-i[\chi(N),re^{i\theta}]\,$.

With a straightforward computation we arrive at $\,[\chi(N),re^{i\theta}]=re^{i\theta}$,
and prove that:
\begin{equation*}
\chi(N)=-i\frac{\partial}{\partial\theta} \;\;.
\end{equation*}

The general solution of (\ref{eq:last}) is $T(\vec{p})=R(r)e^{-i\theta}$,
with $R$ an arbitrary function. If we want the classical Dirac operator
as $\lambda=0$ limit, we are forced to choose $R(r)=-r$, and the final solution is:
\begin{align}
D &=-\rho\left(\begin{array}{cc} 0 & \tfrac{1}{\lambda}\sinh(\lambda E)
+e^{-\lambda E}P\bigl(\tfrac{\lambda}{2}P-i\bigr) \\
\tfrac{1}{\lambda}\sinh(\lambda E)+e^{-\lambda E}P\bigl(
\tfrac{\lambda}{2}P+i\bigr) & 0 \end{array}\right) \nonumber \\
&=\left(\begin{array}{cc} 0 & \tfrac{1}{\lambda}\sinh(i\lambda\partial_0)
+e^{i\lambda\partial_0}\bigl(1+\tfrac{\lambda}{2}\partial_1\bigr)\partial_1 \\
\tfrac{1}{\lambda}\sinh(i\lambda\partial_0)-e^{i\lambda\partial_0}\bigl(1-
\tfrac{\lambda}{2}\partial_1\bigr)\partial_1 & 0 \end{array}\right) \;\;.
\end{align}

This Dirac operator is very similar to the one constructed
in~\cite{NowD,Bib}, but for Euclidean signature instead of
Lorentzian one, and $1+1$ dimension instead of $3+1$.

A deformed Leibniz rule for $D$ comes from the coproduct of $E,P$,
and tells us that commutators with the algebra are not bounded,
due to the presence of the $e^{\lambda E}$ factor. \\
Using equation (2.5) of~\cite{Bib} one can reach the same conclusion for
the Dirac operators in \cite{NowD,Bib}.

From (\ref{eq:Cas}) we derive the following relation:
\begin{equation*}
D^2=\rho\Big(m^2\big(1+\tfrac{\lambda^2 m^2}{4}\big)\Big)\;\;,
\end{equation*}
that is the same as equation (12) in~\cite{NowD}.

\section{Conclusion}\noindent
In quantum mechanics over $\R^n$, it is usual to work in momentum space by
means of Fourier transform. If the physical system under consideration lives in
a curved manifold, with trivial tangent bundle, ``momenta'' are globally
defined as vector fields on the manifold, and in general do not commute. We
have illustrated this situation with an example that is recurrent in physics,
the deSitter space in three dimensions. In such a situation, if we want to work
in ``momentum space'', we need the tools of noncommutative geometry.

In these notes, we have studied the lowest dimensional non--trivial
(i.e.~non--commutative) example, when the manifold $M$ is the connected
component of the group of affine transformations of the real line, and
found that the ``dual'' space is $\kappa$-Minkowski. This is the unique
two-dimensional noncommutative example coming from a Lie group.

We have found a natural representation of $\kappa$-Minkowski on $L^2(M,\mu)$,
with $\mu$ the (left) Haar measure on the group $M$.

Usually this spacetime is studied from an Hopf-algebra point of view,
considering $U(\mathrm{Lie}\,M)$ as the polynomial algebra of coordinates on
some virtual space~\cite{Maj94-Bcp}. We have argued how, using Weyl quantization,
it is possible to define the associated $C^*$-algebra.
The definition of the $C^*$-algebra allows one to study the topology of the space,
following the general philosophical viewpoint that $C^*$-algebra theory may be regarded
as a kind of non-commutative topology. 

In the last section, we have constructed a spectral triple associated to the
cyclic integral of~\cite{DJMTWW}, and an $U_\kappa(iso(2))$-equivariant Dirac operator which
does not satisfies the axioms for a spectral triple.

Apart for its intrinsic interest, $\kappa$-Minkowski is a simple non-trivial
example in which to compare Connes approach to noncommutative geometry with
the Hopf-algebraic one.


\begin{acknowledgments}
\noindent I would like to thank Prof.~G.~Landi for valuable comments and suggestions.
\end{acknowledgments}

\appendix\section{Quantum homogeneous spaces}\label{app}\noindent
Let $G$ be a (compact) topological group that acts on a (compact,
Hausdorff) topological space $M$, $m_0\in M$ a fixed point and $K\subset
G$ the subgroup that leaves $m_0$ invariant. $M$ is called \emph{homogeneous}
if it consists of a single $G$-orbit (every two points can be connected by a
transformation from $G$). It is well known that in this case that the functions
$C(M)$ can be viewed as the subset of $C(G)$ of $K$-invariant functions.

In the noncommutative case, when an Hopf algebra $H$ coacts on an algebra $A$,
we say that $A$ is an \emph{embeddable} space for $H$ if there exists an inclusion
$j:A\hookrightarrow H$ that relates the coaction with the coproduct $\Delta$ of $H$
through one of the formulas:
\begin{equation*}
\Delta\circ j=(id\otimes j)\circ\Delta_L \;\;, \qquad
\Delta\circ j=(j\otimes id)\circ\Delta_R \;\;,
\end{equation*}
depending on which coaction one is considering: a left one $\Delta_L$ or a right
one $\Delta_R$.

Commutative homogeneous spaces $C(M)$ are embeddable in $C(G)$. Then, embeddable
noncommutative spaces provide a possible generalization for the notion of
homogeneous space.




\begin{thebibliography}{9}

\bibitem{Sny47-QST} H.S.~Snyder, \textit{Quantized space--time}, Phys.~Rev.~\textbf{71}
 (1947) 38.

\bibitem{Maj94-Bcp} S.~Majid and H.~Ruegg, \textit{Bicrossproduct structure
 of kappa Poincar\'e group and noncommutative geometry}, Phys.~Lett.
 \textbf{B334} (1994) 348-354.

\bibitem{NowD} A.~Nowicki, E.~Sorace and M.~Tarlini, \textit{The Quantum Deformed Dirac
 Equation from the k-Poincar\`e Algebra}, Phys.~Lett.~\textbf{B302} (1993) 419-422.

\bibitem{Bib} P.N.~Bibikov, \textit{Dirac operator on $\kappa$-Minkowski space bicovariant
 differential calculus and deformed U(1) gauge theory}, J.~Phys.~\textbf{A31} (1998) 6437-6447.

\bibitem{DJMTWW} M.~Dimitrijevic, L.~Jonke, L.~M{\"o}ller, E.~Tsouchnika, J.~Wess
 and M.~Wohlgenannt, \textit{Deformed Field Theory on kappa-spacetime},
 Eur.~Phys.~J.~\textbf{C31} (2003) 129-138.

\bibitem{ConBook} A.~Connes, {\it Noncommutative Geometry}, Academic Press, 1994.

\bibitem{ConRe} A.~Connes, \textit{Noncommutative geometry and reality}, 
 J.~Math.~Phys.~\textbf{36} (1995) 6194-6231.

\bibitem{Con96-Gra} A.~Connes, \textit{Gravity Coupled with Matter and the
 Foundation of Noncommutative Geometry}, Comm.~Math.~Phys.~\textbf{182} (1996)
 155.

\bibitem{Lizzi} J.M.~Gracia--Bondia, F.~Lizzi, G.~Marmo and P.~Vitale,
 \textit{Infinitely many star products to play with}, JHEP \textbf{0204} (2002) 026.

\bibitem{Gay03} V.~Gayral, J.M.~Gracia-Bond{\'\i}a, B.~Iochum, T.~Sch\"ucker and J.C.~Varilly,
 \textit{Moyal Planes are Spectral Triples}, Commun.~Math.~Phys.~\textbf{246} (2004)
 569-623.

\bibitem{Madore} J.~Madore, \textit{The fuzzy sphere}, Class.~Quantum Grav.~\textbf{9}
 (1992) 69-87.

\bibitem{Ful91-RT} W.~Fulton and J.~Harris, \textit{Representation theory},
 Springer-Verlag (1991).

\bibitem{Luk91-qD} J.~Lukierski, H.~Ruegg, A.~Nowicki and V.N.~Tolstoy,
 \textit{q-Deformation of Poincar\'{e} algebra}, Phys.~Lett.~\textbf{B264} (1991) 331.


\bibitem{HA} S.~Majid, \textit{Foundations of quantum group theory}, Cambridge Univ.~Press (2000);

\bibitem{Cel90-E2} E.~Celeghini, R.~Giachetti, E.~Sorace and M.~Tarlini,
 \textit{Three-dimensional quantum groups from contractions of $SU_q(2)$},
 J.~Math.~Phys.~\textbf{31} (1990) 11.

\bibitem{Luk} J.~Lukierski and A.~Nowicki, \textit{Heisenberg Double Description of
 kappa-Poincare Algebra and kappa-deformed Phase Space}, \texttt{q-alg/9702003}.

\bibitem{Now} A.~Nowicki, \textit{Kappa-deformed space-time uncertainty relations},
 \texttt{q-alg/9702004}.

\bibitem{GAC} G.~Amelino-Camelia, J.~Lukierski and A.~Nowicki, \textit{kappa-Deformed
 Covariant Phase Space and Quantum-Gravity Uncertainty Relations}, 
 Phys.~Atom.~Nucl.~\textbf{61} (1998) 1811-1815; Yad.~Fiz.~\textbf{61} (1998) 1925-1929.


\bibitem{Gel} I.M.~Gelfand and M.A.~Naimark, \textit{Unitary representations of
 the group of affine transformations of the straight line}, Dokl.~AN SSSR~\textbf{55}
 (1947) 571--574.

\bibitem{CM} A.~Connes and H.~Moscovici, \textit{The local index formula in
  noncommutative geometry}, Geom.~Funct.~Anal.~\textbf{5} (1995), no.~2, 174-243.

\bibitem{Hig} N.~Higson, \textit{The residue index theorem}, Lecture
  notes for the 2000 Clay Institute symposium on NCG;
  \textit{The local index formula in noncommutative geometry}, Lectures given at
  the School on Algebraic K-Theory and its applications, Trieste 2002.
  Available at the url \texttt{http://www.math.psu.edu/higson/ResearchPapers.html}.

\bibitem{Base} P.~M\'aslanka, \textit{Deformation map and Hermitian representations of
 $\kappa$-Poincar\'e algebra}, J.~Math.~Phys.~\textbf{34} (1993) 6025.
\end{thebibliography}
\end{document}